\documentclass[onecolumn,prd,preprintnumbers,floatfix,article]{revtex4}
\usepackage{dcolumn}
\usepackage{bm}
\usepackage{longtable}
\usepackage{graphicx,epsfig,latexsym,amssymb}
\usepackage{multirow,amsmath,array,booktabs,color}
\usepackage[section]{placeins}
\usepackage{makecell}
\usepackage{array}

\begin{document}
	\title{$J/\psi$ associated production with a bottom quark pair from the Higgs boson decay in next-to-leading order QCD}
	
	\author{Xue-An Pan$^{1}$}
	\author{Zhong-Ming Niu$^{1}$~}
	\author{Mao Song$^{1}$~}
	\author{Yu Zhang$^{2,1}$~}
	\author{Gang Li$^{1}$~}\email{lig2008@mail.ustc.edu.cn}
	\author{Jian-You Guo$^{1}$~}

	\affiliation{$^1$ School of Physics and Optoelectronics Engineering, Anhui University, Hefei, Anhui 230601, People's Republic of China}
	\affiliation{$^2$ Institutes of Physical Science and Information Technology, Anhui University, Hefei, Anhui 230601, People's Republic of China }

	\date{\today }
	%-------------------------------------------------------------------------------------------------------
	\begin{abstract}
In this work, we investigate the next-to-leading order (NLO) QCD correction to $J/\psi$ associated production with a bottom quark pair from the Higgs boson decay within the nonrelativistic QCD framework. From numerical results, {we find that the decay width of process $H \rightarrow b+ J/\psi+\bar{b}$ at leading order (LO) mainly comes from the contribution of the Fock state $^3S^{(8)}_1$, and the NLO QCD corrections significantly enhance the decay width at LO accuracy by about 2 times. At NLO accuracy, the Fock states $^3S^{(8)}_1$ and $^3P^{(8)}_J$ channels give the main contribution, accounting for about $68\%$ and $29\%$ of the total decay width of $J/\psi$ associated production with a bottom quark pair at NLO accuracy from the Higgs boson decay, respectively.
Considering the dominant contribution of color octet (CO) channels at NLO accuracy, the inclusive decay process $H\to b+J/\psi+\bar b + X$ has the potential to be found in future colliders with high energy/luminosity.} The study of $J/\psi$ associated production with a bottom quark pair from the Higgs boson decay is not only useful to study the mechanism of color-octet, but also to assist in the investigation of the coupling for the Higgs boson with the bottom quark.
	\end{abstract}

	\maketitle
	
	\section{Introduction}
As the production and decay processes of heavy quarkonium involve both perturbative and nonperturbative behaviors, it is important for the investigation of perturbative and nonperturbative QCD, which attracts a wide attention by phenomenologist and experimentalist.
The first heavy quarkonium $J/\psi$, composed of charm and anticharm quarks, have been discovered in 1974 \cite{E598:1974sol,SLAC-SP-017:1974ind},
which marks the arrival of heavy quarkonium physics. Subsequently, these processes with various heavy quarkonium production, such as $\psi(2S),\chi_{cJ}, \Upsilon$, and $\chi_{bJ}$, etc., were studied experimentally.
Theoretically, the color-singlet model (CSM)\cite{Einhorn:1975ua,Chang:1979nn,Berger:1980ni}, the color-evaporation model(CEM)\cite{Barger:1979js,Barger:1980mg} and nonrelativistic quantum chromodynamics (NRQCD) \cite{Bodwin:1994jh} are proposed to explain the production and decay of heavy quarkonium. At present, the NRQCD is widely accepted and provides a systematic framework to factorize the quarkonium production and decay into the short-distance coefficients (SDCs) and long-distance matrix elements (LDMEs) by double series in $\alpha_s$ and $v$ expansions. Some phenomenological successes have been achieved in the investigation of heavy quarkonium at the leading order (LO) accuracy within the NRQCD factorization framework, such as it can successfully explain the puzzle of $J/\psi$ and $\psi(2S)$ surplus production at the Tevatron\cite{Braaten:1994vv,Cho:1995vh}, and the prediction of LO NRQCD for the photoproduction process of $J/\psi$ can nicely explain the data taken by the DELPHI Collaboration at LEP2\cite{Todorova-Nova:2001hjh,Klasen:2001cu}.
 However, the predicted results of LO NRQCD about some production processes of heavy quarkonium are inconsistent with the experimental results.
  The prediction with LO NRQCD encounters difficulties in explaining the polarization puzzle of $J/\psi$\cite{Braaten:1999qk,Kniehl:2000nn} and the production of double charmonium at the B factories\cite{Belle:2002tfa,BaBar:2005nic,Belle:2009bxr}, which forces people to consider the contribution of higher-order and other production mechanisms. After considering the contribution of next-to-leading order(NLO) and relativistic corrections, these processes seem to be understandable\cite{Chao:2012iv,Ma:2010yw,Zhang:2005cha,Zhang:2006ay,Gong:2007db,He:2007te}, and people find that the higher-order corrections may play an important role in the production process of heavy quarkonium\cite{Zhou:2016zbn,Mao:2015ada,Ma:2014svb}. In the past two decades, many works have considered the contribution of higher-order corrections to heavy quarkonium and achieved remarkable successes in phenomenologically(see the review\cite{Lansberg:2019adr} and its references). However, NRQCD still faces challenges, such as the universality of LDMEs\cite{Ma:2010yw,Zhang:2009ym}, the kinematic distribution of $J/\psi+W^{\pm}$ measured by ATLAS\cite{ATLAS:2019jzd,Baranov:2021gil}, and  the polarization problem of $\chi_{c1}$ and $\chi_{c2}$\cite{Chapon:2020heu} have not been fully explained. In order to study the production mechanism of heavy quarkonium, we should study more the production process of heavy quarkonium at different colliders.

 The Higgs boson, as the last discovered fundamental particle in the standard model (SM), plays a considerable role in testing the SM and searching for new physics beyond SM, it has aroused profound interest from phenomenology and experiment.{At present, the total and differential rates predicted of all the possible production and decay channels of the Higgs boson in the SM so far are consistent with those measurements of experiments in the theoretical and experimental uncertainties\cite{ATLAS:2019nkf,CMS:2018uag}, but the couplings of the Higgs boson to the electrons and lighter quarks of the first and second generations are yet to be established either phenomenology or experiments. This indicates that the properties of the Higgs boson have not been fully understood, and need to be further studied. The process of the Higgs boson decay to heavy quarkonium has been widely studied as an important decay channel for investigating the properties of the Higgs boson and the production mechanism of heavy quarkonium, such as the processes of $H \rightarrow J/\psi(\Upsilon) +\gamma$\cite{Bodwin:2013gca,Chao:2016usd,Brambilla:2019fmu,Bodwin:2016edd,Bodwin:2017wdu},$H \rightarrow J/\psi(\Upsilon) +Z$\cite{Modak:2014ywa}, $H \rightarrow J/\psi +J/\psi$\cite{Kartvelishvili:2008tz} and the Higgs boson decay to $\chi_{b}$\cite{Sun:2019cxx} have been investigated. Furthermore, the decay processes of the Higgs decay to $B_c$\cite{Jiang:2015pah,Liao:2018nab} and doubly heavy baryons\cite{Niu:2019xuq} have been studied. Experimentally, in recent years, people have proposed to build colliders with high energy/luminosity, which can collect a large number of Higgs boson events.} For example, Circular Electron-Positron Collider (CEPC), which is considered as a Higgs factory, can collect $1.10 \times 10^6$ Higgs events per year when the center-of-mass energy $\sqrt{s} = 250$ GeV\cite{Jiang:2015pah,Ahmad1}. HE-LHC(HL-LHC) is an upgraded collider of Large Hadron Collider (LHC)\cite{Todesco,LHCnew}, with the characteristics of high energy (high luminosity), and can produce reach up $6.0 \times 10^8$($1.65\times 10^8$) the Higgs boson events per year\cite{LHCnew,Liao:2018nab}. These colliders provide an ideal platform for the investigation of precise Higgs physics, especially for the process of the Higgs boson decay to quarkonium. Although the rate of the Higgs boson decay to heavy quarkonium is slight, due to its clean decay channel, these rare decay processes of the Higgs boson have the potential to be observed on the excellent experimental platform mentioned above, which will help us to further study the properties of the Higgs boson and the production mechanism of quarkonium.

In Ref.\cite{Qiao:1998kv}, the decay process of the Higgs boson, $H \rightarrow b + J/\psi +\bar{b}$, has been studied at LO accuracy by Qiao \emph{et al.} for the first time, and they find that the process might play an important role in the Higgs boson decay to the charmonium production process, further considering the final states of this process has high detection efficiency in experiment, thus the decay process $H \rightarrow b + J/\psi +\bar{b}$ may be worthy of further study. In consideration of the fact that the higher-order QCD corrections has a significant contribution to the production process of heavy quarkonium. In this work, we calculate NLO QCD corrections to the inclusive decay process $H \rightarrow b + J/\psi +\bar{b} + X$ within the NRQCD framework by applying the covariant projection method\cite{Petrelli:1997ge}.
While considering the contribution of NLO corrections, we find the decay width for the process $H \rightarrow b + J/\psi + \bar{b} $ at LO is significantly enhanced by the NLO QCD corrections, and it has the potential to be detected in future colliders with high energy or high luminosity. This will not only be an auxiliary study for the coupling of the Higgs boson with the bottom quark, but will also help us to further investigate the production mechanism of heavy quarkonium. The remainder of this paper is organized as follows. In Sec. II, we introduce the details of the calculation framework for the process $H \rightarrow b + J/\psi + \bar{b}$ at NLO accuracy.  The input parameters and numerical results are presented in Sec. III. A brief summary and discussion are collected in Sec. IV.

\section{calculation framework}
\begin{figure*}[!htbp]
		\hspace*{1cm}
		\includegraphics[scale=1]{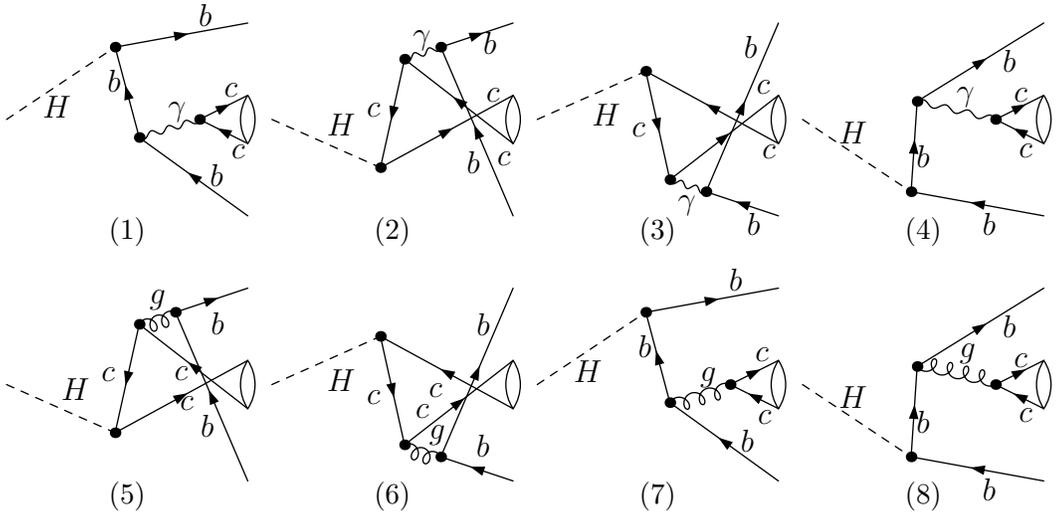}\newline
		\caption{The Feynman diagrams for the decay process $H \rightarrow b + J/\psi + \bar{b}$ at $\alpha^2$ and $\alpha_s^2$ orders, respectively.}
		\label{born}
	\end{figure*}

In this section, we introduce the details of the calculation for the $J/\psi$ associated production with a bottom quark pair from the Higgs boson decay up to NLO accuracy within the NRQCD factorization framework\cite{Petrelli:1997ge}. The decay process of the Higgs boson, $H \rightarrow b+J/\psi + \bar{b}$ is denoted as

\begin{eqnarray}
H(p_1) \rightarrow b(p_2) + J/\psi(p_3) + \bar{b}(p_4),
\label{process}
\end{eqnarray}
with $p_1,p_2,p_3$, and $p_4$ are the momentum of the Higgs bosons and its decay products, respectively, and the corresponding Feynman diagrams are shown in Fig.\ref{born}. The Feynman diagrams of the decay process $H \rightarrow b+J/\psi + \bar{b}$ at $\alpha^2$ and $\alpha_s^2$ orders are presented in Fig.\ref{born}(1)-(4) and Fig.\ref{born}(5)-(8), respectively, in which the contribution of the former to the decay process $H \rightarrow b+J/\psi + \bar{b}$ is much smaller than that of the latter, the decay process $H \rightarrow b+J/\psi + \bar{b}$ at $\alpha^2$ order can be ignored. In our calculation, the decay process $H \rightarrow b+J/\psi + \bar{b}$ at LO is considered as the decay process at $\alpha_s^2$ order, which the corresponding Feynman diagrams are shown in Fig.\ref{born}(5)-(8).

The decay width of the process $H \rightarrow b+J/\psi + \bar{b}$ at LO accuracy can be expressed as
\begin{eqnarray}
\Gamma_{\rm LO}(^3S^{(8)}_1)=\hat{\Gamma}(^3S^{(8)}_1)\left<\mathcal{O}^{J/\psi}(^3S^{(8)}_1)\right>.
\label{LO}
\end{eqnarray}
The $\left<\mathcal{O}^{J/\psi}(^3S^{(8)}_1)\right>$ is the LDMEs, which can be obtained from experiments, and $\hat{\Gamma}(^3S^{(8)}_1)$ denotes the short-distance decay width of Fock state $c\bar{c}(^3S^{(8)}_1)$ channel, which can be written as
\begin{eqnarray}
\hat{\Gamma}(^3S^{(8)}_1) = \frac{1}{2m_H N_{\rm pol} N_{\rm col}}\int \sum\left|M_{\rm LO}\right|^2d\Phi_3,
\label{SDCs}
\end{eqnarray}
where $m_H$ is the mass of the Higgs boson, $N_{\rm pol}(N_{\rm col})$ denotes the polarization (color) quantum number of Fock state $c\bar{c}(^3S^{(8)}_1)$, and the summation is taken over the color and spin states of the initial and final state. $d\Phi_3$ is 3-body differential phase space:
\begin{eqnarray}
d\Phi_3=(2\pi)^D\delta^D\left(p_1-\sum_{f=2}^{4}p_f\right)\prod^{4}_{f=2}\frac{d^{D-1}{\bm p}_f}{(2\pi)^{D-1}2E_f}.
\end{eqnarray}

With these formulas, the LO decay width for the process $H \rightarrow b+J/\psi + \bar{b}$ can be calculated directly. From our calculations, we find that the contribution from the Feynman diagrams with the Higgs-bottom quark direct coupling is much greater than that from the Feynman diagrams with the Higgs-charm quark direct coupling. Therefore, in this work, we only consider the contributions from the Feynman diagrams of Higgs-bottom quark direct coupling for the decay process $H \rightarrow b+J/\psi + \bar{b}$, and the corresponding Feynman diagrams are shown in Fig.\ref{born}(7)-(8). When ignoring the mass of the bottom quark(except for in the Higgs-bottom quark coupling vertex), we can obtain the analytical differential decay width for the process $H \rightarrow b + J/\psi + \bar{b}$ at LO accuracy, which is the same as Ref.\cite{Qiao:1998kv}. In our investigation, the mass of the bottom quark is retained.

\begin{figure}[!htbp]
		\hspace*{1cm}
		\includegraphics[scale=1]{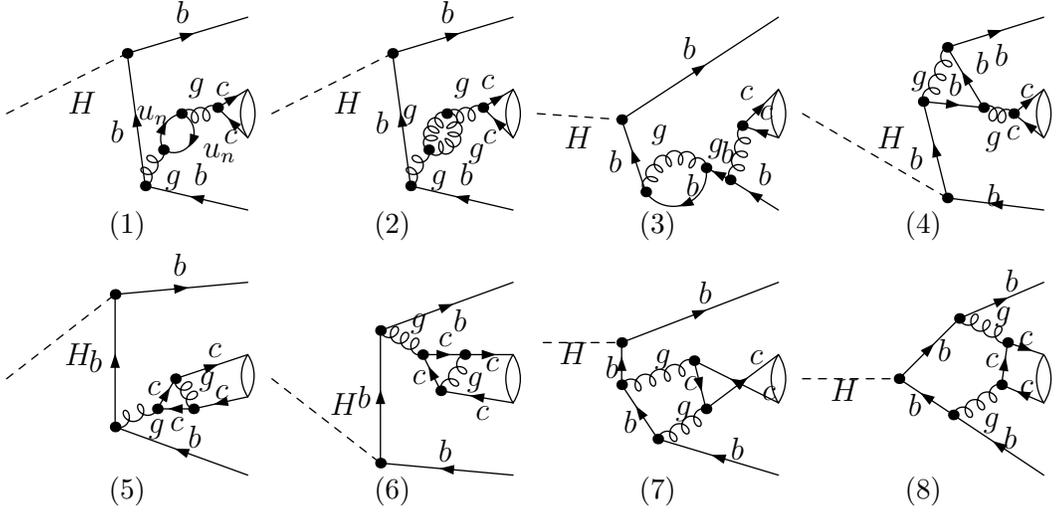}\newline
		\caption{Some representative one-loop Feynman diagrams for the decay process $H \rightarrow b + J/\psi + \bar{b}$.}
		\label{virtual}
	\end{figure}

When calculating the NLO QCD corrections to the decay process $H \rightarrow b + J/\psi + \bar{b}$, we should consider the contribution of both the virtual corrections and real gluon radiation corrections. In this work, the contribution of virtual correction only comes from the Fock state $c\bar{c}(^3S_1^{(8)})$, and some representative one-loop Feynman diagrams are shown in Fig.\ref{virtual}.
The contribution of virtual correction can be calculated by the following formula:
\begin{eqnarray}
\Delta\Gamma_{\rm virtual}(^3S^{(8)}_1)=\Delta\hat{\Gamma}_{\rm virtual}(^3S^{(8)}_1)\left<\mathcal{O}^{J/\psi}(^3S^{(8)}_1)\right>,
\label{v1}
\end{eqnarray}
 and

\begin{eqnarray}
\Delta\hat{\Gamma}_{\rm virtual}(^3S^{(8)}_1) = \frac{1}{2m_H N_{\rm pol} N_{\rm col}} \int \sum 2{\rm Re}(M^{*}_{\rm LO}M_{\rm virtual}) d\Phi_3,
\label{SDCs1}
\end{eqnarray}
with $M_{\rm virtual}$ referring to the amplitudes corresponding to the one-loop Feynman diagrams.

\begin{figure}[!htbp]
		\hspace*{1cm}
		\includegraphics[scale=1]{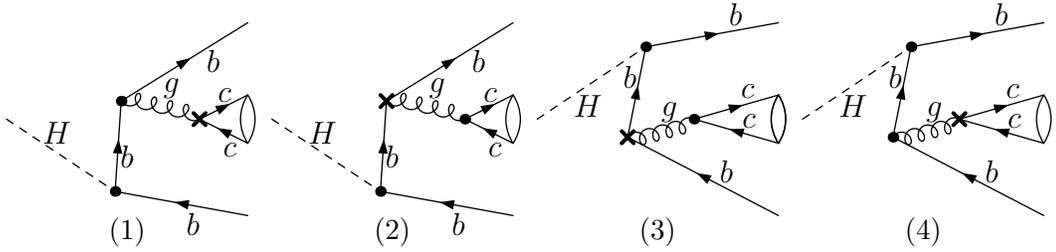}\newline
		\caption{Some representative counterterms Feynman diagrams for the decay process $H \rightarrow b + J/\psi + \bar{b}$, in which the dashed, straight, cycles and bold cross line refers to the Higgs boson, bottom or charm (anti)quarks, gluon, and the NLO QCD counterterms for $q\bar{q}g(q=c,b)$ vertices, respectively.}
		\label{counterterm}
	\end{figure}

When calculating the virtual corrections of the Fock states $c\bar{c}(^3S_1^{(8)})$ in the one-loop Feynman diagrams, there will be ultraviolet (UV) divergences, infrared (IR) divergences, and Coulomb singularities.  In order to cancel off the UV and IR divergences, we adopt the dimensional regularization(DR) scheme with $D=4-2\epsilon$, and some representative counterterm Feynman diagrams are present in Fig.\ref{counterterm}. The quark wave functions and the strong coupling constant are renormalized by on-mass-shell (OS) and the modified minimal subtraction ($\overline{\rm MS}$), respectively. The renormalization constants $Z_{m_i}$, $Z_{2i}$, $Z_3$, and $Z_g$ correspond to quark mass $m_i$, quark field $\psi_i$, gluon field $A^a_{\mu}$, and QCD gauge coupling $g_s$, respectively. These renormalization constants can be expressed as follows:
\begin{eqnarray}
\delta Z^{\rm OS}_{m_i} &=& -3C_F \frac{\alpha_s}{4\pi}\left[\frac{1}{\epsilon_{\rm UV}}-\gamma_E+\ln\frac{4\pi \mu^2_r}{m_i^2}+\frac{4}{3}\right],	\notag \\
\delta Z^{\rm OS}_{2_i} &=& -C_F \frac{\alpha_s}{4\pi}\left[\frac{1}{\epsilon_{\rm UV}} + \frac{2}{\epsilon_{\rm IR}}-3\gamma_E+3\ln\frac{4\pi \mu^2_r}{m_i^2}+4\right], \notag	\\
\delta Z^{\rm OS}_{3} &=& \frac{\alpha_s}{4\pi}\left[(5-\frac{2n_f}{3})(\frac{1}{\epsilon_{\rm UV}}-\frac{1}{\epsilon_{\rm IR}})-\frac{4}{3}T_F\left(2(\frac{1}{\epsilon_{\rm UV}}-\gamma_E)+ \ln\frac{4\pi \mu^2_r}{m_c^2} + \ln\frac{4\pi \mu^2_r}{m_b^2}\right)\right], \notag \\
\delta Z_g^{\rm \overline{MS}} &=& -\frac{\beta_0}{2}\frac{\alpha_s}{4\pi}\left[\frac{1}{\epsilon_{\rm UV}}-\gamma_E+\ln(4\pi)\right],
\end{eqnarray}
where $i$ refers to the bottom and charm quark, and the $1/\epsilon_{\rm UV}$ ($1/\epsilon_{\rm IR}$) denotes the UV(IR) divergence. $\gamma_E$ and $\mu_r$ are the Euler's constant and renormalization scales, respectively. In our calculation, the number of light quark flavors($n_f$) is considered to be 3. $\beta_0=\frac{11}{3}C_A-\frac{4}{3}T_F(n_f+2)$ is the one-loop coefficient of the QCD beta function with $C_A=3$, $T_F=\frac{1}{2}$, and $C_F=\frac{4}{3}$.

In our calculation, we chose the ${\rm \overline{MS}}$ mass of the bottom quark as the bottom quark mass in the Yukawa coupling, and the renormalization constant for the corresponding ${\rm \overline{MS}}$ mass of the bottom quark as:
\begin{eqnarray}
(\frac{\delta \bar{m}_b}{\bar{m}_b})_{{\rm \overline{MS}}}=-3C_F \frac{\alpha_s}{4\pi}\left[\frac{1}{\epsilon_{\rm UV}}-\gamma_E+\ln4\pi\right].
\end{eqnarray}
Except for the bottom quark mass in the Yukawa coupling, we keep the pole mass of the bottom quark everywhere else.

After considering the contribution of counterterms, the UV divergence of the virtual corrections in the one-loop Feynman diagrams can be completely canceled, however, there are still some IR divergence and Coulomb singularities that have not been eliminated. We find that only Fig.\ref{virtual}(5) and Fig.\ref{virtual}(6) induce Coulomb singularities, which can be canceled by considering renormalization of the operator $<\mathcal{O}^{J/\psi}[^3S^{(8)}_1]>$, and the remaining IR divergences can be removed by taking into account the contribution of the Fock state $^3S^{(8)}_1$ channel in the real gluon radiation process.

\begin{figure}[!htbp]
		\hspace*{1cm}
		\includegraphics[scale=1]{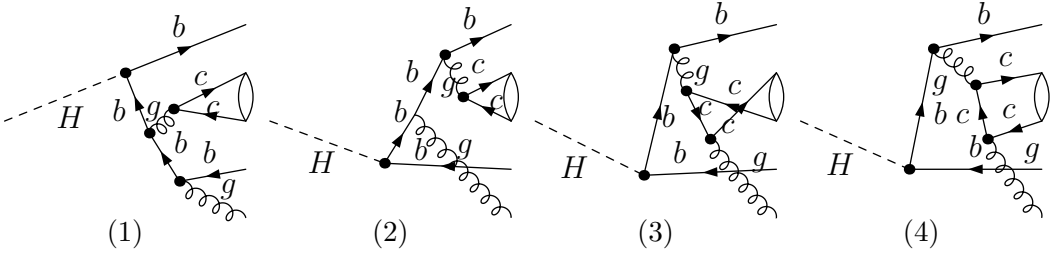}\newline
		\caption{The part Feynman diagrams of the real gluon radiation corrections for the decay process $H \rightarrow b + J/\psi + \bar{b}$}
		\label{real}
	\end{figure}

The process of real gluon radiation corrections can be written as
\begin{eqnarray}
H(p_1) \rightarrow b(p_2) + J/\psi(p_3) + \bar{b}(p_4) + g(p_5),
\label{process2}
\end{eqnarray}
where $g(p_5)$ denotes the real gluon($p_5$ represents the momentum of the real gluon), and the part Feynman diagrams of the real gluon radiation corrections for the decay process $H \rightarrow b + J/\psi + \bar{b}$ are drawn in Fig.\ref{real}.

When considering the contribution of the real gluon radiation process, we should include the contribution of Fock states $c\bar{c}(^1S_0^{(8)})$, $c\bar{c}(^3S_1^{(8)})$, and $c\bar{c}(^3P_J^{(8)})$ channels. The real gluon radiation process with the Fock state $^1S^{(8)}_0$ is free of divergence and can be calculated by using a Monte Carlo technique in four dimensions \cite{Lepage:1977sw}.
When we calculate the contribution of the Fock states $^3S^{(8)}_1$ and $^3P^{(8)}_J$ channels in the real gluon radiation process, soft singularities will arise but no collinear IR singularity exists since the bottom and charm quarks are massive. In order to isolate these IR singularities, we adopt the phase space slicing (PSS) method\cite{Giele:1991vf} to deal with the real gluon radiation process for Fock states $^3S^{(8)}_1$ and $^3P^{(8)}_J$. By introducing a small cutoff $\delta_s$, the phase space of the decay process $H \rightarrow b + J/\psi + \bar{b} + g$ is separated into two regions: the soft region with $E_5 < \delta_s m_H/2$, and the hard region with $E_5 \geq \delta_s m_H/2$, where $E_5$ is the energy of the real gluon.

 The decay width of the Fock state $^3S^{(8)}_1$ channel in the real gluon radiation process can be expressed as the summation of the contribution over the two phase space regions:
\begin{eqnarray}
\Delta \Gamma_{{\rm real}}({^3S_1^{(8)}}) = \Delta \Gamma_{{\rm hard}}({^3S_1^{(8)}}) + \Delta \Gamma_{{\rm soft}}({^3S_1^{(8)}}).
\end{eqnarray}
The $\Delta \Gamma_{{\rm hard}}({^3S_1^{(8)}})$ refers to the contribution of the hard region with Fock state $^3S_1^{(8)}$ channel in the decay process $H \rightarrow b + J/\psi + \bar{b} + g$, which is finite and also can be integrated by using a Monte Carlo technique in four dimensions \cite{Lepage:1977sw}, and the $\Delta \Gamma_{{\rm soft}}({^3S_1^{(8)}})$ denotes the contribution from the soft region in the decay channel, which contain the soft singularities. These soft singularities can completely cancel each other with the remaining IR singularities in the virtual corrections after considering the contribution of counterterms. The expression of $\Delta \Gamma_{{\rm soft}}({^3S_1^{(8)}})$ can be obtained as Refs.\cite{Beenakker:2001rj,Beenakker:2002nc,Catani:1996jh,Catani:2002hc}
\begin{eqnarray}
\Delta \Gamma_{{\rm soft}}({^3S_1^{(8)}}) = -\frac{\alpha_s}{2\pi}\left[-\frac{3}{2}\left(g(p_2,p_c)+g(p_4,p_c)\right)+\frac{1}{6}g(p_2,p_4)\right]\Gamma_{\rm LO}({^3S_1^{(8)}}),
\end{eqnarray}
with $p_c=p_{\bar{c}}=\frac{p_3}{2}$, and $g(p_i,p_j)$ is the soft integral as Ref.\cite{Beenakker:2001rj}
\begin{eqnarray}
g(p_i,p_j) = \frac{(2\pi \mu_r)^{2\epsilon}}{2\pi}\int_{E_5 \le \delta_s m_H/2} \frac{d^{D-1}\boldsymbol{p_5}}{E_5}\left[\frac{2(p_i\cdot p_j)}{(p_i\cdot p_5)(p_j \cdot p_5)}-\frac{p_i^2}{(p_i\cdot p_5)^2}-\frac{p_j^2}{(p_j\cdot p_5)^2}\right].
\end{eqnarray}

As for the contribution of the real gluon radiation process with Fock state $^3P_J^{(8)}$, we use the same strategy as for the Fock state $^3S_1^{(8)}$ channel in the decay process $H \rightarrow b + J/\psi + \bar{b} + g$ to calculate the decay width, which can be expressed as

\begin{eqnarray}
\Delta\Gamma_{{\rm real}}({^3P^{(8)}_J}) =  \Delta\Gamma_{{\rm hard}}({^3P^{(8)}_J}) + \Delta\Gamma_{{\rm soft}}({^3P^{(8)}_J}),
\end{eqnarray}
where the $\Delta\Gamma_{{\rm hard}}({^3P^{(8)}_J})$ is finite and can be integrated with four dimensions by using the Monte Carlo method, and the $\Delta\Gamma_{{\rm soft}}({^3P^{(8)}_J})$ can be obtained by adopting the method of Ref.\cite{Petrelli:1997ge}

\begin{eqnarray}
\Delta\Gamma_{{\rm soft}}({^3P^{(8)}_J}) &=& -\left(\frac{1}{\epsilon}-2\ln\delta_s + \frac{1}{\beta}\ln\frac{1+\beta}{1-\beta}\right)\frac{4\alpha_sB_F}{3\pi m^2_c} \notag \\
                                         && \times \frac{\Gamma(1-\epsilon)}{\Gamma(1-2\epsilon)}\left(\frac{4\pi\mu^2_r}{\hat{s}}\right)^{\epsilon} \hat{\Gamma}(^3S^{(8)}_1) \notag \\
                                         &&\times \left<\mathcal{O}^{J/\psi}[^3P^{(8)}_J]\right> ,
\end{eqnarray}
with $\beta =\sqrt{1-4m_c^2/E_3^2}$, $E_3$ is the energy of $J/\psi$, and $B_F=\frac{N^2_c-4}{4N_c}=\frac{5}{12}$ with $N_c=3$.

After the renormalization of the operator $<\mathcal{O}^{J/\psi}[^3S^{(8)}_1]>$ by taking into account the NRQCD NLO corrections, the Coulomb singularity in virtual corrections and the soft IR divergences of the Fock state $^3P^{(8)}_J$ channel in the real gluon radiation process will be canceled out, and we adopt the same method as Ref.\cite{Klasen:2004tz} to renormalize the operator $<\mathcal{O}^{J/\psi}[^3S^{(8)}_1]>$ as

\begin{eqnarray}
\left<\mathcal{O}^{J/\psi}\left[^3S^{(8)}_1\right]\right>_{\rm Born} &=& \left<\mathcal{O}^{J/\psi}\left[^3S^{(8)}_1\right]\right>_{r}(\mu_{\rm \Lambda})\left[1-\left(C_F-\frac{C_A}{2}\right)\frac{\pi\alpha_s}{2v}\right] \notag \\
&& + \frac{4\alpha_s}{3\pi m_c^2}\left(\frac{4\pi\mu^2_r}{\mu^2_{\rm \Lambda}}\right)^{\epsilon}\exp(-\epsilon \gamma_E)\frac{1}{\epsilon} \notag \\
&& \times \sum_{J=0}^{2}\left(C_F\left<\mathcal{O}^{J/\psi}[^3P^{(1)}_J]\right>+B_F\left<\mathcal{O}^{J/\psi}[^3P^{(8)}_J]\right>\right),
\label{reborn}
\end{eqnarray}
 where $2v$ is the small relative velocity between $c$ and $\bar{c}$ in the meson rest frame, $\mu_{\rm \Lambda}$ denotes the NRQCD scales.

Finally, adding the contribution of all the above parts, we can get the finite total decay width of the process $H \rightarrow b+J/\psi + \bar{b}$ up to NLO accuracy as
\begin{eqnarray}
\Gamma_{\rm NLO} &=&  \Gamma_{{\rm NLO}}({^3S^{(8)}_1}) + \Delta \Gamma_{{\rm real}}({^1S^{(8)}_0}) + \Delta \Gamma_{{\rm real}}({^3P^{(8)}_J}) \notag \\
    &=& \Gamma_{{\rm LO}}({^3S^{(8)}_1}) + \Delta \Gamma_{{\rm virtual}}({^3S^{(8)}_1}) +\Delta \Gamma_{{\rm real}}({^3S^{(8)}_1}) +  \Delta \Gamma_{{\rm real}}({^1S^{(8)}_0}) + \Delta \Gamma_{{\rm real}}({^3P^{(8)}_J}).
\end{eqnarray}

\begin{figure}[!htbp]
		\hspace*{1cm}
		\includegraphics[scale=0.45]{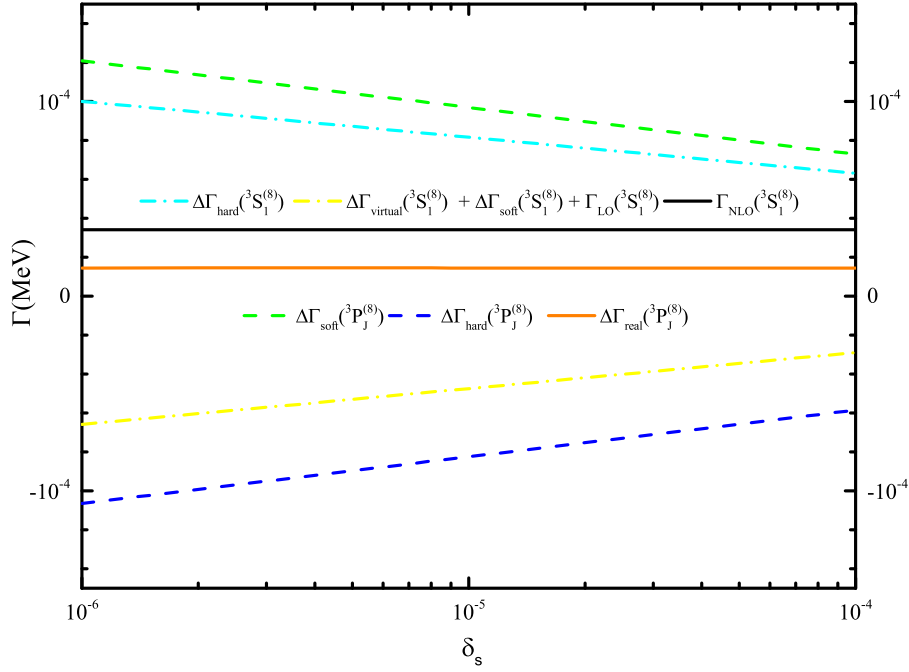}\newline
		\caption{Verifications of the independence on the small cutoff $\delta_s$.}
		\label{delta_s}
      \end{figure}

Theoretically, the total decay width of the process $H \rightarrow b+J/\psi + \bar{b}$ at NLO accuracy is independent of arbitrary small cutoff $\delta_s$. In order to test the independence of cutoff $\delta_s$ for the contribution of the Fock states $^3S^{(8)}_1$ and $^3P^{(8)}_J$ CO channels, we present the decay width of the Fock states $^3S^{(8)}_1$ and $^3P^{(8)}_J$ CO channels with the change of cutoff $\delta_s$ in Fig. \ref{delta_s}. From the figure, we can find that the $\Delta\Gamma_{{\rm NLO}}({^3S^{(8)}_1})$ and $\Delta\Gamma_{{\rm real}}({^3P^{(8)}_J})$ remain almost unchanged when the value of the cutoff $\delta_s$ varies from $10^{-6}$ to $10^{-4}$.  Such stable numerical results prove that our results are independent of arbitrary small cutoff $\delta_s$. The value of cutoff $\delta_s$ both for the Fock states $^3S^{(8)}_1$ and $^3P^{(8)}_J$ are taken as $\delta_s=10^{-4}$ in our calculation.

In this paper, we use the FeynArts package\cite{Hahn:2000kx} to generate the Feynman diagrams and Feynman amplitudes for the decay process $H \rightarrow b + J/\psi +\bar{b}$ at LO and NLO, and we further reduce Feynman amplitudes using FeynCalc\cite{Shtabovenko:2016sxi}. In particular, we use Apart\cite{Feng:2012iq} and FIRE\cite{Smirnov:2008iw} packages to reduce the Feynman amplitudes corresponding to one-loop Feynman diagrams. The numerical results are given by the LoopTools and FormCalc packages\cite{Hahn:1998yk}.

	\section{numerical results}

In this section, we present the numerical results for the decay process $H \rightarrow b + J/\psi + \bar{b} $ up to NLO accuracy. For the numerical calculation, the masses of the Higgs, $Z$, $W^{\pm}$ boson, various quarks, and $J/\psi$ meson are taken as

 \begin{eqnarray}
 &&m_H =125.1 \ {\rm GeV}, m_Z=91.1876\ {\rm GeV}, m_W=80.399\ {\rm GeV}, \notag \\
 &&m_q=0(q=u,d,s), m_t=172 \ {\rm GeV}, m_b=4.9 \ {\rm GeV}, m_c = 1.5\ {\rm GeV}, M_{J/\psi} =2 m_c,
\end{eqnarray}
and the fine structure constant, factorization, renormalization, and NRQCD scales are chosen as $\alpha=1/137.036$, $\mu_f=\mu_r=m_H$, and $\mu_{\rm \Lambda} = m_c$, respectively. The ${\rm \overline{MS}}$ mass of the bottom quark can be obtained by the expressions as\cite{Dawson:2003kb,Sun:2007ihy}:
\begin{eqnarray}
\bar{m}_b(\mu_r)_{1l}=m_b\left[\frac{\alpha_s(\mu_r)}{\alpha_s(m_b)}\right]^{c_0/b_0}
\label{1-loopmass}
\end{eqnarray}
\begin{eqnarray}
\bar{m}_b(\mu_r)_{2l}=m_b\left[\frac{\alpha_s(\mu_r)}{\alpha_s(m_b)}\right]^{c_0/b_0}\left[1+\frac{c_0}{b_0}(c_1-b_1)[\alpha_s(\mu_r)-\alpha_s(m_b)]\right]\left(1-\frac{4}{3}\frac{\alpha_s(m_b)}{\pi}\right),
\label{2-loopmass}
\end{eqnarray}
where
\begin{eqnarray}
b_0&=&\frac{1}{4\pi}\left(\frac{11}{3}N-\frac{2}{3}n_{lf}\right), c_0=\frac{1}{\pi}, \notag \\
b_1&=&\frac{1}{2\pi}\frac{51N-19n_{lf}}{11N-2n_{lf}}, c_1=\frac{1}{72\pi}(101N-10n_{lf}).
\label{parameter}
\end{eqnarray}
The $\bar{m}_b(\mu_r)_{1l}$, and $\bar{m}_b(\mu_r)_{2l}$ refer to the $\overline{\rm MS}$ mass of the bottom quark $\bar{m}_b(\mu_r)$ in the Yukawa coupling at the LO and NLO, respectively. $N=3$ and $n_{lf} = 5$ is the number of colors and the number of light flavors, respectively.

The LDMEs of $J/\psi$ is taken from Ref.\cite{Butenschoen:2012qh} as
\begin{eqnarray}
\label{LDMEs}
\left\langle\mathcal{O}^{J / \psi}\left[{ }^{1} S_{0}^{(8)}\right]\right\rangle &=& (3.04\pm{0.35}) \times 10^{-2}\ \mathrm{GeV}^{3}, \notag \\
\left\langle\mathcal{O}^{J / \psi}\left[{ }^{3} P_{0}^{(8)}\right]\right\rangle &=& (-9.08\pm{1.61}) \times 10^{-3}\ \mathrm{GeV}^{5},  \\
\left\langle\mathcal{O}^{J / \psi}\left[{ }^{3} S_{1}^{(8)}\right]\right\rangle &=& (1.68\pm {0.46}) \times 10^{-3}\ \mathrm{GeV}^{3}, \notag
\end{eqnarray}
and we can obtain the values of $\langle\mathcal{O}^{J / \psi}[{ }^{3} P_{1}^{(8)}]\rangle$ and $\langle\mathcal{O}^{J / \psi}[{ }^{3} P_{2}^{(8)}]\rangle$ by the relation:
\begin{eqnarray}
\left\langle\mathcal{O}^{J / \psi}\left[{ }^{3} P_{J}^{(8)}\right]\right\rangle =(2J+1)\left\langle\mathcal{O}^{J / \psi}\left[{ }^{3} P_{0}^{(8)}\right]\right\rangle.
\end{eqnarray}

\begin{table}[!htbp]
\caption{The decay width (in units of $ {\rm eV}$) for the process $H \rightarrow b+ J/\psi + \bar{b}$ at the LO and NLO including different CO channels. }
\renewcommand\arraystretch{3}
\setlength{\tabcolsep}{5mm}
\begin{tabular}{c|c|c|c|c|c}
\hline\hline
 &  $\Delta \Gamma_{{\rm real}}({^1S^{(8)}_0})$ & $\Delta \Gamma_{{\rm real}}({^3P^{(8)}_J})$ & $ \Gamma_{{\rm NLO}}({^3S^{(8)}_1})$ & $\Gamma_{\rm NLO}$ & $\Gamma_{\rm LO}({^3S^{(8)}_1})$  \\ \hline
 Decay width($ \ {\rm eV}$ ) & $2.0 $ & $14.4 $ & $34.1 $ & $50.5 $& $15.8 $
 \\
  \hline \hline
\end{tabular}
\label{Table1}
\end{table}

	In Table \ref{Table1}, we present the decay width for the process $H \rightarrow b+ J/\psi + \bar{b}$ at LO and NLO including different CO channels. From the Table, we can find that the decay width for the process $H \rightarrow b+J/\psi+\bar{b}$ at LO is significantly enhanced by NLO QCD corrections, and the contribution of NLO QCD corrections is about 2 times that of LO. The Fock state $^3S^{(8)}_1$ plays a major role in the process $H \rightarrow b+J/\psi+\bar{b}$ at NLO accuracy, from which the contribution can account for about $68\%$ of the total decay width at NLO, and the contribution of Fock state $^3P^{(8)}_J$ is also considerable, which accounts for about $29\%$.
Considering the total decay width of the Higgs boson is 4.1 MeV \cite{LHCHiggsCrossSectionWorkingGroup:2016ypw}, we can obtain the branching ratio of the decay process $H \rightarrow b+ J/\psi+\bar{b}$ as
\begin{eqnarray}
{\rm Br}(H \rightarrow b+ J/\psi+\bar{b}) = {1.23 \times 10^{-5}},
\end{eqnarray}	
it has the potential to be found in future colliders with high energy/luminosity, { which is not only useful to study the mechanism of the color-octet, but also to assist in the investigation of the coupling for the Higgs boson with the bottom quark.}

	\begin{figure}[!htbp]
		\hspace*{1cm}
		\includegraphics[scale=0.30]{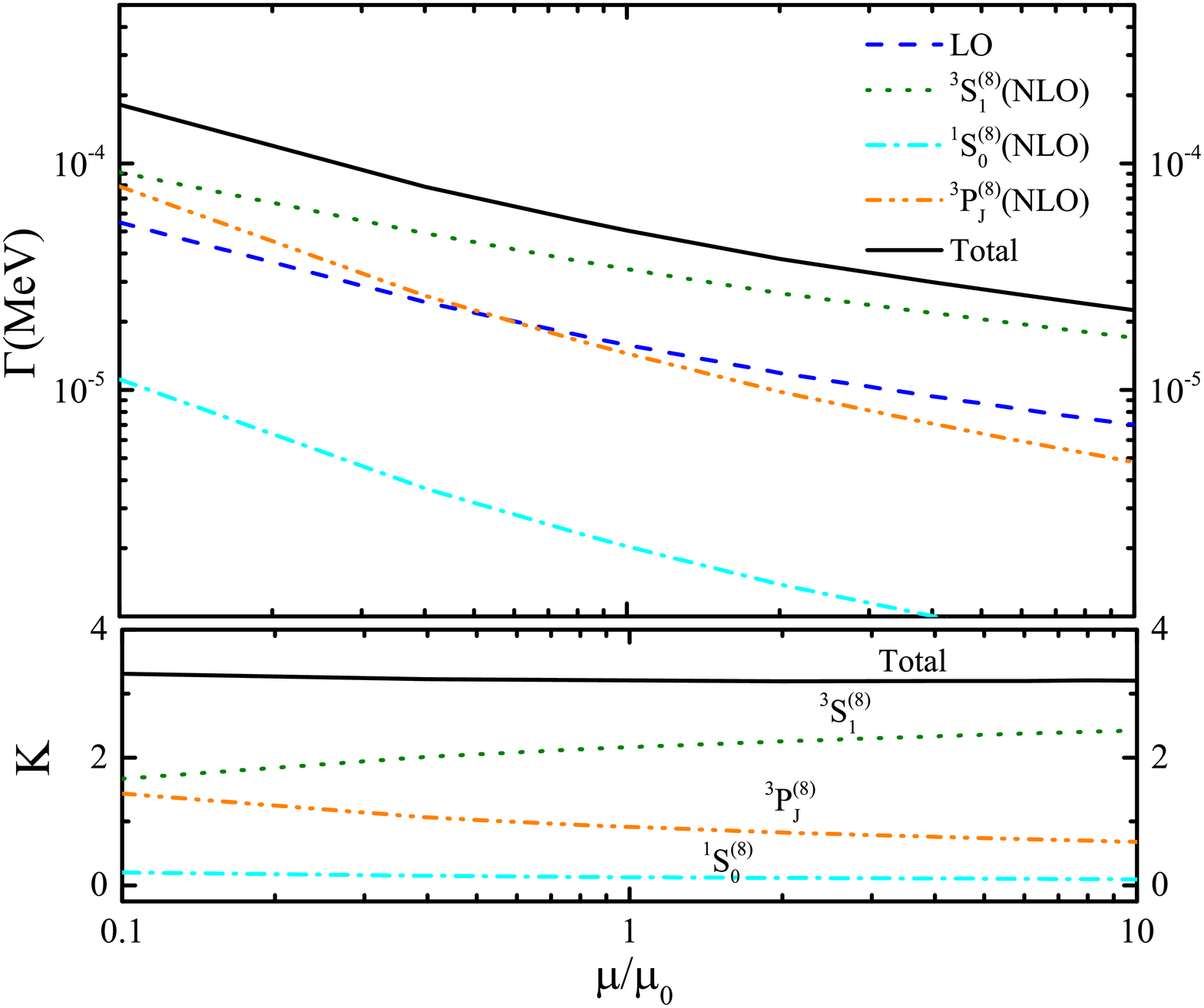}\newline
		\caption{The $\mu$ dependence of the LO decay width, the NLO decay channel including the contribution of different CO channels, and the corresponding $K$ factors ($K(\mu) = \Gamma_{\rm NLO}(\mu)/\Gamma_{\rm LO}(\mu)$) for the decay process $H \rightarrow b + J/\psi +\bar{b}$. Here we assume $\mu=\mu_f=\mu_r$ and define $\mu_0=m_H$.}
		\label{scale}
      \end{figure}

We further investigate the dependence of decay width for the process $H \rightarrow b+J/\psi +\bar{b}$ at LO and the NLO accuracy on the renormalization and factorization scale. Fig. \ref{scale} shows the curves of the LO decay width, NLO decay width including the contribution of different CO channels, and the $K$ factors ($K(\mu) = \Gamma_{\rm NLO}(\mu)/\Gamma_{\rm LO}(\mu)$) of the corresponding CO channel with the varies of scale $\mu$ for the decay process $H \rightarrow b + J/\psi +\bar{b}$. In our calculation, we assume $\mu=\mu_f=\mu_r$ and define $\mu_0=m_H$. As illustrated in this figure, we can find that the $K$ factor of total decay width is consistently above 3 in the whole of plotted region of scale $\mu$ and slightly decrease with the increase of scale $\mu$. At NLO accuracy, the Fock states $^3S^{(8)}_1$ CO channel also give the main contribution to the decay process $H \rightarrow b+J/\psi +\bar{b}$, and we find that the scale $\mu$ dependence has slightly improved. When the scale $\mu$ from $0.1\mu_0$ to $10\mu_0$, the corresponding $K$ factor is about 2.0 and slightly increases with the rise of scale $\mu$. As for the contribution of Fock state $^3P^{(8)}_J$ CO channel also cannot be ignored and decreases with the increase of scale $\mu$.

 As we know, the uncertainty of the results for the heavy quarkonium production process is greatly affected by the values of LDMEs. We also consider the uncertainty for the inclusive decay process $H \rightarrow b + J/\psi +\bar{b}+X$ due to the uncertainty of LDMEs values. After considering the range of ${ }^{1} S_{0}^{(8)}$, ${ }^{3} P_{0}^{(8)}$, and ${ }^{3} S_{1}^{(8)}$ LDMEs values, we can obtain the decay width of the corresponding channels are $(2.0\pm 0.2)\ {\rm eV}$,$(14.4 \pm 2.6)\ {\rm eV}$, and $(34.1\pm 9.3) \ {\rm eV}$, respectively. And the total decay width of the process $H \rightarrow b + J/\psi +\bar{b}$ at the NLO accuracy varies between $76\% - 124\%$ compared with the results with the center values of LDMEs.

	\section{summary and discussion}
In this paper, we have investigated the $J/\psi$ associated production with a bottom quark pair from the Higgs boson decay at NLO accuracy within the NRQCD framework. We give the numerical results of decay width for the process $H \rightarrow b+ J/\psi+\bar{b}$ at LO and NLO, respectively, and present the dependence of the decay width of different CO channels on the factorization and renormalization scale. From numerical results, {we find that the decay width of the process $H \rightarrow b+ J/\psi+\bar{b}$ at LO dominantly comes from Fock state $^3S^{(8)}_1$, and the decay width is significantly enhanced by the NLO QCD corrections.}  At NLO accuracy, the Fock state $^3S^{(8)}_1$ channel also gives the main contribution, which account for about $68\%$ of the total decay width, and the contribution of Fock state $^3P^{(8)}_J$ channel is considerable, which account for about $29\%$. Experimentally, the $J/\psi$ meson and bottom quark have high detection efficiency, and considering that the future high energy/luminosity collider provides an ideal platform for the study of the Higgs boson, the  inclusive decay process $H \rightarrow b + J/\psi +\bar{b} + X$ has the potential to be found in future colliders with high energy/luminosity. {In conclusion, the process of $J/\psi$ associated production with a bottom quark pair from the Higgs boson decay deserves further study, it is not only useful to study the mechanism of the color-octet, but also to assist in the investigation of the coupling for the Higgs boson with the bottom quark.}

	\section*{Acknowledgments}
	This work was supported in part by the National Natural
	Science Foundation of China (Grants No. 11805001, No. 11935001, and No. 11875070), and the Natural Science Foundation of Anhui Province (Grants No. 2108085MA20).

\end{document}